# Micromechanical Experimental and Numerical Studies of Collagen Fibers Failure in Arterial Tissue

Xiaochang Leng[1], Yingchao Yang[3], Xiaomin Deng[1], Susan M. Lessner[2,4], Michael A. Sutton[1,2], Tarek Shazly[1,2]

[1] College of Engineering and Computing, Department of Mechanical Engineering
University of South Carolina
Columbia, SC 29208

[2] College of Engineering and Computing, Biomedical Engineering Program
University of South Carolina
Columbia, SC 29208

[3] Department of Mechanical Engineering
Maine University
Orono, ME 04469

[4] School of medicine, Department of Cell Biology & Anatomy
University of South Carolina
Columbia, SC 29208

**ABSTRACT**

Arterial tissue failures lead to a number of clinical conditions that develop rapidly and unpredictably in vivo. Structural components and their interfacial mechanical strength of arterial tissue play a critical role in the process of arterail delamination. Therefore, there is a pressing need to understand the micromechanical mechanisms of arterial delamination. The objective of this study was to investigate various failure mechanisms (e.g. failure of collagen fibers) responsible for arterial interfacial delamination. In-situ tensile tests of fibers were performed on a micro-tester in the scanning electron microscope. A 3D unit cell model containing an individual fiber bridging two arterial tissue layers was constructed. An exponential cohesive zone model (CZM) was used to assess the stiffening and softening mechanical behaviors of collagen fiber bundles between the two arterial layers. An anisotropic constitutive model was implemented for characterizing the mechanical properties of the amorphous matrix which includes the fibrous cap and the underlying plaque tissue and a nonlinear elastic model was adopted for characterizing the mechanical properties of the fibers. The CZM and elastic parameter values of fibers were identified through an inverse boundary value approach that matches the load-displacement curves from simulation predictions of tensile test of collagen fibers with experimental measurements. The identified parameter values were then used as input in the 3D unit cell model, through which micromechanical factors affecting the resultant traction-separation relation for the unit cell were investigated via a parametric analysis. Results of the parametric analysis showed the applicability of the 3D unit cell model approach for evaluating the micromechanical mechanisms of arterial tissue failure processes.

# 1. Introduction

Plaque rupture is manifested as a complicated mechanical failure phenomenon, often in the form of fibrous cap dissection or medial delamination from the underlying arterial wall, leading to life-threatening clinical consequences such as heart disease and stroke. The American Heart Association estimated the medical cost for coronary heart disease and stroke will reach to $276 billion in 2030 [1].

The arterial tissue failure has been speculated to correlate with the age-related extracellular matrix remodeling within the aortic wall [2-4]. This remodeling process is a common potential cause for arterial stiffening, which is related to the development of strokes and heart failure [5, 6]. Meanwhile, the age-related aortic remodeling occurs at multiple length scales: at the macroscale, aging aortic diameters and arch length increased significantly [7-10]; at the microscale, the micromechanical remodeling in the aging artery includes: (i) changes in the amount of fibrillar collagen, elastin [11-14] and subtype of collagen fiber (type I and type III) [13, 15-17]; (ii) collagen fibers degradation [18-21].

The mechanical properties of compositions of arterial tissue at the microscale are essential for determining the failure process in the arterial layers. The measured local energy release rate of the interface inside arterial wall varies over a wide range [22, 23]. This suggests that the local energy release rate depend on local plaque composition, such as collagen fibers (colorless strands 1 to 100 $\mu m$ thick [24-26]), which play a crucial role in determining the mechanical behavior of the aortic wall [27, 28]. These observations have led us to believe that damage accumulation before delamination, oscillation of the force-displacement curve during delamination, and the distribution of critical energy release rates reflect differences in the mechanisms of collagen fibers failure behavior (collagen fibers bridging) at the arterial tissue interface [22].

A method based on atomic force microscopy (AFM) was performed on single collagen fibrils isolated from collagenous materials to quantify Young's modulus [29, 30], bending and shear modulus [31], and another method using nanoindentation by AFM was conducted on single collagen fibrils to acquire the reduced modulus which provides new insight into collagen structure. The yield stress and strain at failure of collagen fibrils from sea cucumber were obtained through uniaxial tensile testing [32]. Kato et al. [33] performed tensile tests on collagen fibers from rat tail tendons to obtain the tensile strengths. Miyazaki and Hayashi [34] acquired the modulus, tensile strength and strain at failure through uniaxial tensile test of collagen fibers from rat tail tendons. For the modulus, tensile strength and strain at failure of collagen fascicles from the rabbit patellar tendon, those material properties were quantified from tensile testing [35]. However, these studies do not provide the mechanical properties such as modulus and adhesive strength bundles of collagen fibers from arterial wall.

We have successfully developed a mechanical modeling approach for atherosclerotic plaque rupture in the form of plaque delamination (dissection) along the plaque-media interface and fibrous cap delamination along fibrous cap-underlying plaque tissue interface [23, 36, 37]. However, the arterial tissue is a complex, laminated structure which contains several types of fiber-reinforced elements such as fiber bridges extending perpendicular to the interface between plaque and media or fibrous cap and underlying plaque tissue. Although the macroscale cohesive zone model (CZM) approach is useful for understanding of the overall arterial delamination behavior, it does not give the descriptions of the micromechanical-physical basis for the arterial tissue delamination process. Furthermore, the structural elements in the artery are individual cells and extracellular matrix fibers. Most drug treatments are effective to alter the relative amounts, mechanical properties, or interactions between these individual components [38]. Therefore, in order to assess the effects of drug use or medical intervention on the arterial tissue failure, we need to understand how the mechanical properties and organization of the individual components at the microscale, especially the collagen fibers, contribute to the adhesive strength of interfaces in the arterial layers.

The aim of this study was to gain a mechanical understanding of the plaque rupture phenomenon at the microscopic scale. Firstly, a uniaxial tensile test on bundles of collagen fibers from porcine arterial wall was performed to acquire the elastic modulus, tensile stress and strain at failure. Secondly, the adhesive strength of interface across fibers was obtained through best fitting of the load-displacement curve from the simulation predictions with the experimental measurements. Finally, these parameter values were used as input for a micromechanical model of a plaque-arterial wall system, which was developed based on experimental observations and the cohesive zone model approach. The failure mechanisms at the microscopic scale (such as collagen fibers breakage) were incorporated to develop a three-dimensional unit cell model, enabling the characterization of the cohesive traction-separation relation and the factors at the micromechanical scale affecting this relation which plays an important role for the understanding of micromechanical mechanism in plaque rupture.

## 2. Materials and methods

*Specimen preparation*

All tissue handling protocols were approved by the Institutional Animal Care and Use Committee at the University of South Carolina. The porcine abdominal aorta was harvested from the kidneys of the male Landrace Pigs (age 8-12 months, mass 60-70 kg). All kidneys were acquired immediately after animal sacrifice and placed on ice for transport to laboratory. Upon kidney arrival, the abdominal aorta was isolated from the surrounding tissue, washed in phosphate buffered saline (PBS), dissected free of perivascular tissue. Strips of 20 mm ×10 mm were cut using surgical scissors (Fig. 1a). Scalpel was used to make a 2 mm long cut in the middle of the strip and tear it apart, releasing the collagen fibers embedded in the tissue. The separated surfaces obtained from this process are similar to those from the arterial tissue dissection and delamination processes. Hence, the shape and diameter of collagen fibers obtained are the same as the fibers along the surfaces from the delamination or dissection process. A part of teared surface is shown in Fig. 1b, where many bundles of collagen fibers distributed randomly and connected together. Six specimens were harvested at the teared surface, these specimens are consisted with bundles of collagen fibers twisted together.

*Uniaxial tensile testing of collagen fibers*

Both ends of fibers were clamped onto a GatanTM Deben micro-tester (Gatan, Inc., Pleasanton, CA) equipped with a 2 N load cell (Fig. 1c). The fibers together with micro-tester was placed into a scanning electron microscope (SEM: FEI Quanta 400) for in-situ tensile tests. The loading speed was prescribed as 1 mm/min. The whole process of fibers uniaxial tensile test was recorded by a video to reveal the corresponding deformation and fracture processes.

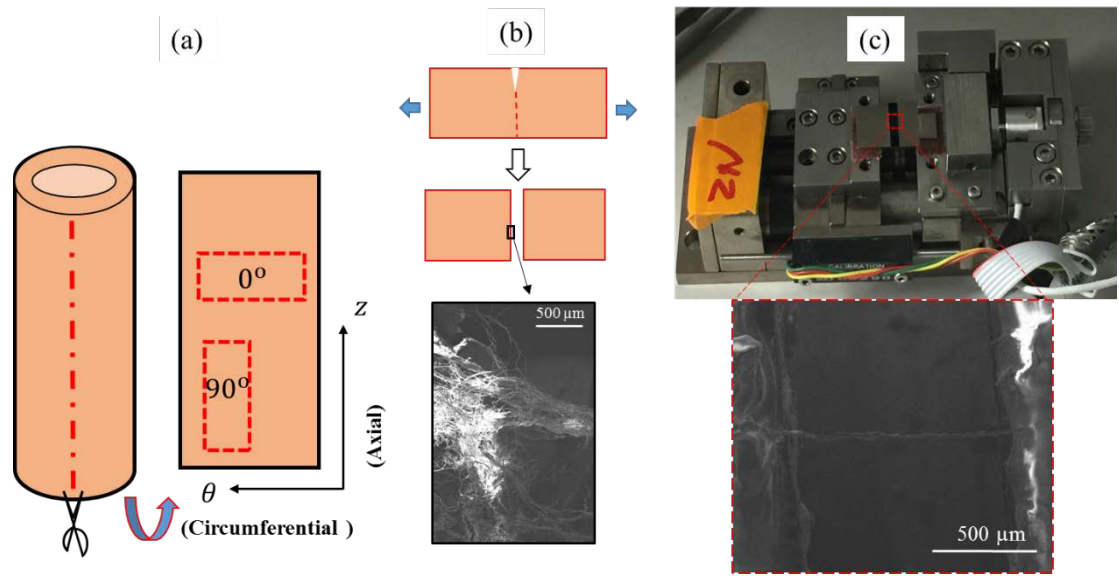

**Fig. 1** Schematic of set-up for tensile testing in SEM. (a) A radial cut was made on the porcine abdominal aorta and strips oriented at the angle of $0^o$ and $90^o$ with respect to the circumferential vessel axis were obtained; (b) a strip was teared into two parts and the collagen fibers was harvested at the teared surface; (c) experimental setup of collagen fibers uniaxial tensile test.

## 3. Theoretical Framework

### 3.1 Bulk material

The Holzapfel-Gasser-Ogden (HGO) model assumes that the mean orientation of collagen fibers lies along the circumferential direction of arterial wall [39, 40], The anisotropic hyperelastic potential $\Psi$ for an arterial layer can be represented

$$\Psi = \frac{1}{D}\left(\frac{J^2-1}{2} - \ln J\right) + \frac{\mu}{2}(\bar{I}_1 - 3) + \frac{k_1}{k_2}\left[e^{k_2[\kappa \bar{I}_1 + (1-3\kappa)\bar{I}_4 - 1]^2} - 1\right] \quad (8\text{-}1)$$

where $\frac{1}{D}$ is analogous to the bulk modulus of the material; $J = \det(F)$ is the determination of deformation gradient $F$; μ is the neo-Hookean parameter; the constitutive parameter $k_1$ is related to the relative stiffness of fibers, which is identified from mechanical tests of tissue; $k_2$ is dimensionless parameter; the parameter $\kappa$ is the dispersion parameter, which characterizes the dispersion of the two families of fibers along the two mean distributed directions, and $0 \leq \kappa \leq 1/3$; the two family collagen fibers are parallel to each other when $\kappa = 0$, whereas the fibers distribute isotropically when $\kappa = 1/3$; γ represents the angle between the mean fiber orientation of one family of fibers and the circumferential direction of the aorta.

### 3.2 Interface damage model

The free energy potential ($\varphi$) per unit (undeformed) area of an exponential CZM model can be written as follows [41],

$$\varphi = e\sigma_c \delta_c \left[1 - \left(1 + \frac{\delta}{\delta_c}\right) \exp\left(-\frac{\delta}{\delta_c}\right)\right] \quad (8\text{-}2)$$

where $e = \exp(1) \approx 2.71828$ and $\sigma_c$ denotes the cohesive strength of the material; $\delta_c$ is the maximum effective displacement when $t = \sigma_c$ and $\delta$ is the effective displacement jump across the cohesive surfaces.

Considering the loading conditions, the cohesive traction is expressed as:

$$t = \frac{\partial \varphi}{\partial \delta} = e\sigma_c \frac{\delta}{\delta_c} exp(-\frac{\delta}{\delta_c}) \qquad (8\text{-}3)$$

The effective displacement jump is given by:

$$\delta = \sqrt{\beta^2 \delta_s^2 + \delta_n^2} \qquad (8\text{-}4)$$

where $\delta_n$ represents the normal displacement jump; $\delta_s$ denotes the sliding displacement across the cohesive surfaces which is in the form,

$$\delta_s = \sqrt{\delta_{s1}^2 + \delta_{s2}^2} \qquad (8\text{-}5)$$

where $\delta_{s1}$ and $\delta_{s2}$ denote components of the sliding displacement ($\delta_s$) of the cohesive interface; $\beta$ is a scalar parameter assigns different weights to the opening displacement ($\delta_n$) and sliding displacement ($\delta_s$).

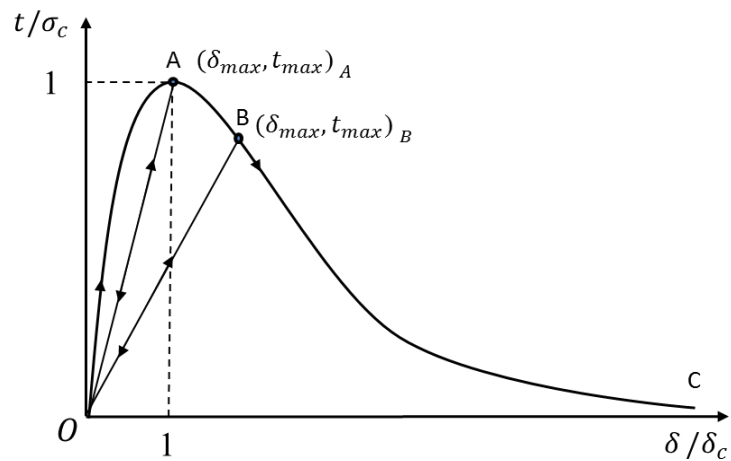

**Fig. 2.** Exponential irreversible cohesive model represents by normalized effective traction ($t$) and effective displacement ($\delta$).

The energy release rate represents the area below the traction-separation curve in Fig. 2 is follows as:

$$G_c = e\sigma_c\delta_c \quad (8\text{-}6)$$

In order to describe the irreversible mechanism, we introduce a damage parameter, representing the damage of the cohesive surfaces during the loading and unloading processes, which defined as,

$$d = \frac{\varphi(\delta_{max})}{G_c} \quad (8\text{-}7)$$

The damage parameter d ranges from 0 to 1, corresponding to a state with no damage of the cohesive surface and a fully separation of the cohesive interface, respectively. Figure 2 illustrates the loading and unloading processes of the exponential cohesive traction-separation law, the damage will initiate and accumulate when the effective traction t is larger than zero. The traction-separation relation will follow line AO or BO when unloading at point A or B because permanent damage occurs at point A and B. The exponential CZM was implemented in the current study via a user UEL subroutine in ABAQUS [42, 43].

## 4. Numerical Implementation

In order to study the mechanism of collagen fibers failure at a micro scale, a 3D unit cell model was implemented in the numerical simulation. Firstly, the CZM parameter values of the fiber-fiber interface were identified via fitting the simulation predicted with experimental measured mechanical response (Fig. 3a and 3b). Then, the CZM parameter values were used as input data for the parametric studies using the micromechanical model (Fig. 3c).

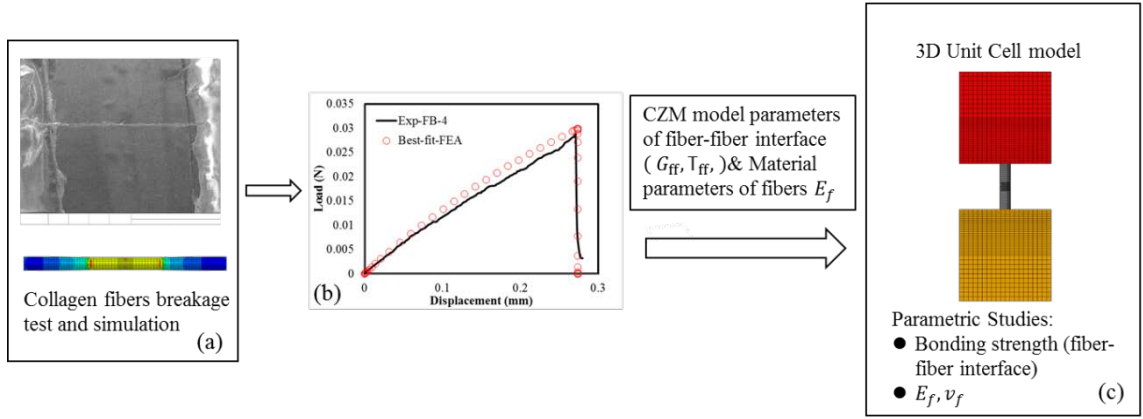

**Fig. 3.** Workflow of the numerical implementation of micromechanical model of the fibers breakage. (a) SEM image of collagen fibers in the uniaxial tensile test and FE model; (b) best fitting of load vs. load-point displacement curve between the simulation predictions and experimental measurements; (c) Schematic illustration of the 3D unit cell model.

## 4.1 Material properties identification process

The material constants for the interfacial strength and critical energy release rate of interface across fibers (fiber-fiber interface) as well as the material properties of collagen fibers were obtained via fitting simulation predictions with experimental measurements of the load vs. load-point displacement curves.

### 4.1.1 Material properties of fiber-fiber interface and collagen fibers

*Geometrical modeling*

The geometry of the collagen fibers was reconstructed from the images obtained during the tensile testing (Fig. 4a). The length and diameter of the specimen were measured directly from the experimental images (Fig. 4a). For simplification, the shape of the cross-section was assumed to be a circular section [44].

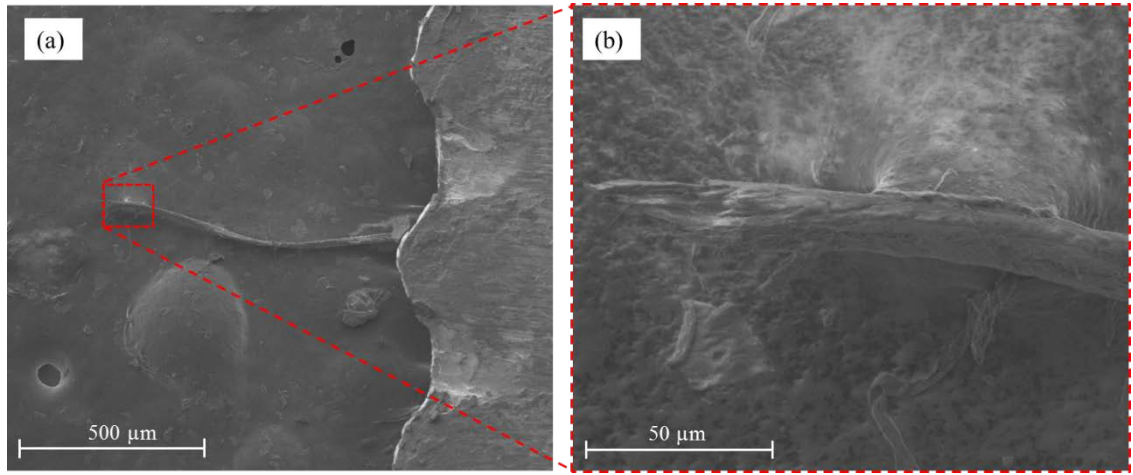

**Fig. 4.** (a) SEM images of fibers after the tensile test and (b) the zoom-in image of the tip of the collagen fibers.

*Meshing*

The geometrical model of the fibers was meshed with eight-node brick elements (C3D8). The zero thickness eight-node 3D user-defined elements were placed across the cross-section perpendicular to the longitudinal axis where the fibers breakage occurred in the experiment.

*Boundary conditions*

In line with the experiments, the left end of the fibers was fixed, the load applied on the right end with loading rate 1.0 mm/min. Except for the fixed part of the model, all other surfaces of FE model were set to a traction-free boundary condition.

## 5. Results

### 5.1 Mechanical properties of collagen fibers

The equations for calculating the Cauchy stress and strain as follows,

$$\sigma = \frac{F}{A_0}\left(1 + \frac{\Delta L}{L_0}\right), \quad \varepsilon = \ln\left(1 + \frac{\Delta L}{L_0}\right) \tag{8}$$

where $\sigma$ and $\varepsilon$ are the Cauchy stress and strain, respectively; $F$ is the resultant load in the tensile testing; $A_0$ and $L_0$ represent the initial cross-section and length of the fibers, respectively; $\Delta L$ denotes the increment of length during the fibers tensile test. The Cauchy stress-strain curves of all specimens were shown in Fig. 5. The failure strain and ultimate tensile strength were 23.00 ± 7.30 % and 81.75 ± 75.14 MPa (Table 1).

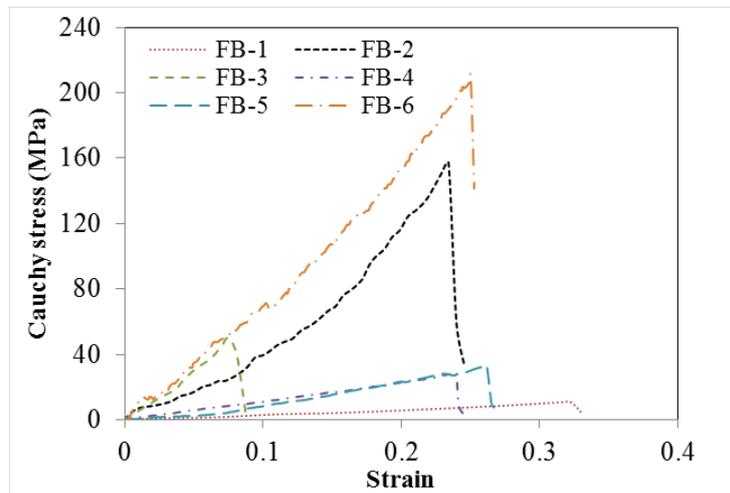

**Fig. 5.** Cauchy stress-strain curves of the samples in the uniaxial tensile tests.

**Table 1.** Diameter, strain at fracture and ultimate strength of the samples.

|  | FB-1 | FB-2 | FB-3 | FB-4 | FB-5 | FB-6 | Mean | S.D. |
|---|---|---|---|---|---|---|---|---|
| Diameter, D (μm) | 28.6 | 11.1 | 16.8 | 39.5 | 32.3 | 12.4 | 23.5 | 10.7 |
| Strain at failure (%) | 32 | 23 | 8 | 24 | 26 | 25 | 23.00 | 7.30 |
| Ultimate Strength (MPa) | 10.81 | 156.88 | 50.29 | 28.34 | 32.35 | 211.82 | 81.75 | 75.14 |

### 5.2 CZM parameter values of fiber-fiber interface

The CZM parameter values for the interface across the fibers were obtained through a numerical identification procedure that matches simulation predictions of the load vs. load-point displacement curve with the experimental measurements [23, 36, 45]. The tangential modulus and ultimate tensile strength obtained from the fibers tensile tests were used as input as the modulus $E$ of fibers and interfacial strength $\sigma_c$ of fiber-fiber interface in the parameters identification procedure. The initial guess of critical energy release rate $G_c$ was chosen according to the critical energy release rate of fibrous cap delamination tests [23]. The value of Poisson's ratio $\nu$ of fibers was taken to be 0.3 [46].

In the numerical identification process, the modulus, Poisson's ratio of fibers and the CZM parameter values of interface across fibers were considered "acceptable" when the root mean square error satisfies

$$f = \frac{\sqrt{\frac{\chi^2}{N-M}}}{F_{avg}} < 0.005 \text{ , with } \chi^2 = \sum_{i=1}^{N}\left[\left(F_{exp} - F_{sim}\right)_i^2\right] \tag{50}$$

where $F_{sim}$ and $F_{exp}$ are the simulation predicted and the experimentally measured resultant forces; $F_{avg}$ is the sum of all experimentally measured forces divided by the number of data points; $N$ is the number of data points on the load-displacement curve that were used in the parameter value identification procedure; and $M$ is the number of parameters whose values were determined from the identification procedure.

A set of modulus and Poisson's ratio of fibers and the CZM parameter values of interface across fibers is shown in Table 2. It is noted that the values for K and $\lambda$ were assumed equal to 1 N/mm$^3$ and 1. These values are reasonable since the K value is sufficiently large that artificial compliance from the cohesive interface can be prevented and the values for parameter $\lambda$ is reasonable for the mode I fracture process of collagen fibers breakage process [23, 36].

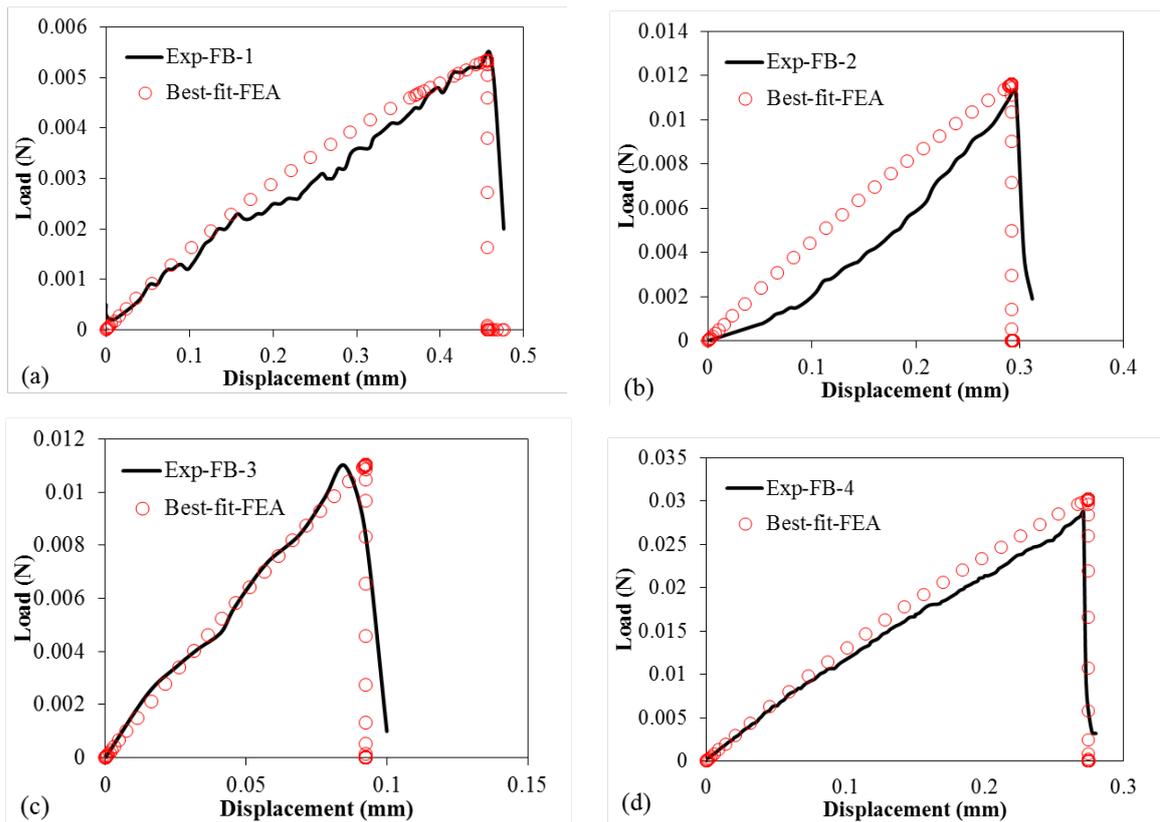

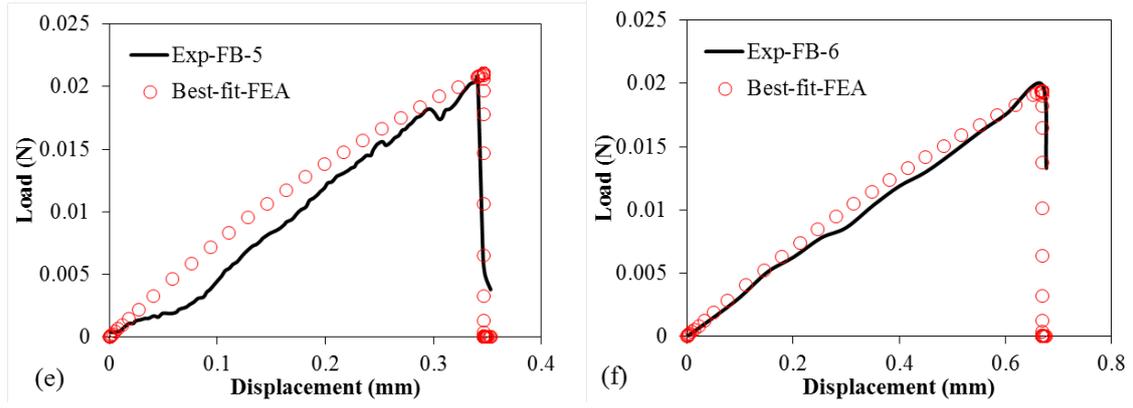

**Fig. 6.** The comparison between numerically predictions and experimental measurements of the load-displacement curves of the samples.

Table 2. Best-fit parameters for the exponential CZM of the collagen fibers.

|  | FB-1 | FB-2 | FB-3 | FB-4 | FB-5 | FB-6 | Mean | S.D. |
|---|---|---|---|---|---|---|---|---|
| Modulus, E (MPa) | 28.02 | 850 | 534.38 | 118.08 | 105.14 | 781.28 | 402.82 | 334.33 |
| $G_c$ (N/mm) | 0.005 | 0.32 | 0.21 | 0.1 | 0.15 | 0.47 | 0.209 | 0.151 |
| $\sigma_c$ (MPa) | 11.32 | 160.82 | 55.75 | 28.34 | 29.39 | 211.82 | 82.91 | 75.71 |
| Residual, $f$ | 0.0003 | 0.0023 | 0.0009 | 0.0047 | 0.0037 | 0.0008 | 0.0021 | 0.0016 |

# 6. Parametric studies

A micromechanical model was proposed to characterize the arterial delamination mechanics at the fibrous cap-underlying plaque tissue interface in terms of the mechanical properties and geometry of fibrous components including bridging fibers. A 3D unit cell containing a set of fibers between two arterial tissue layers was considered.

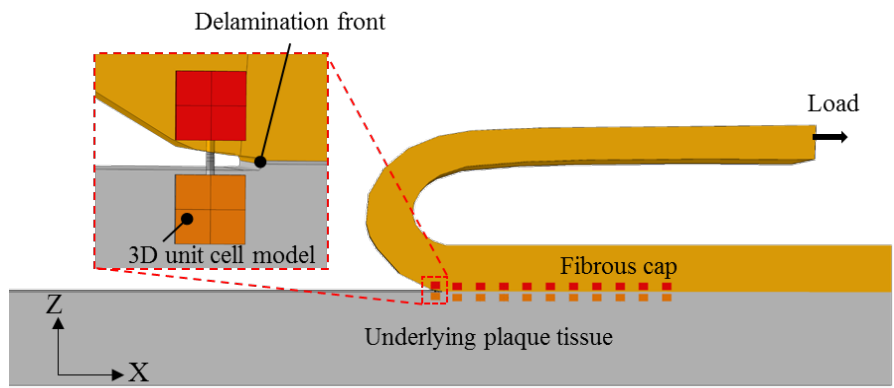

**Fig. 7.** Schematic representation of 3D unit cell model.

The top and bottom arterial tissue layers were modeled as hyperelastic anisotropic materials (HGO model) and the fibers were treated as a linear elastic material. In order to investigate the factors affecting the traction-separation response of delamination process at the micro scale, the parametric studies based on the 3D unit cell model was implemented, which considering: (1) the bonding strength of interface across the fibers; (2) variations in the fibers' stress-strain behavior; (3) initial gap of the interface.

## 6.1  3D unit cell model for the arterial tissue delamination

*Geometrical modeling*

The geometry model was shown in Fig. 8, which was constructed according to the mean values of diameter of fibers in table 1.

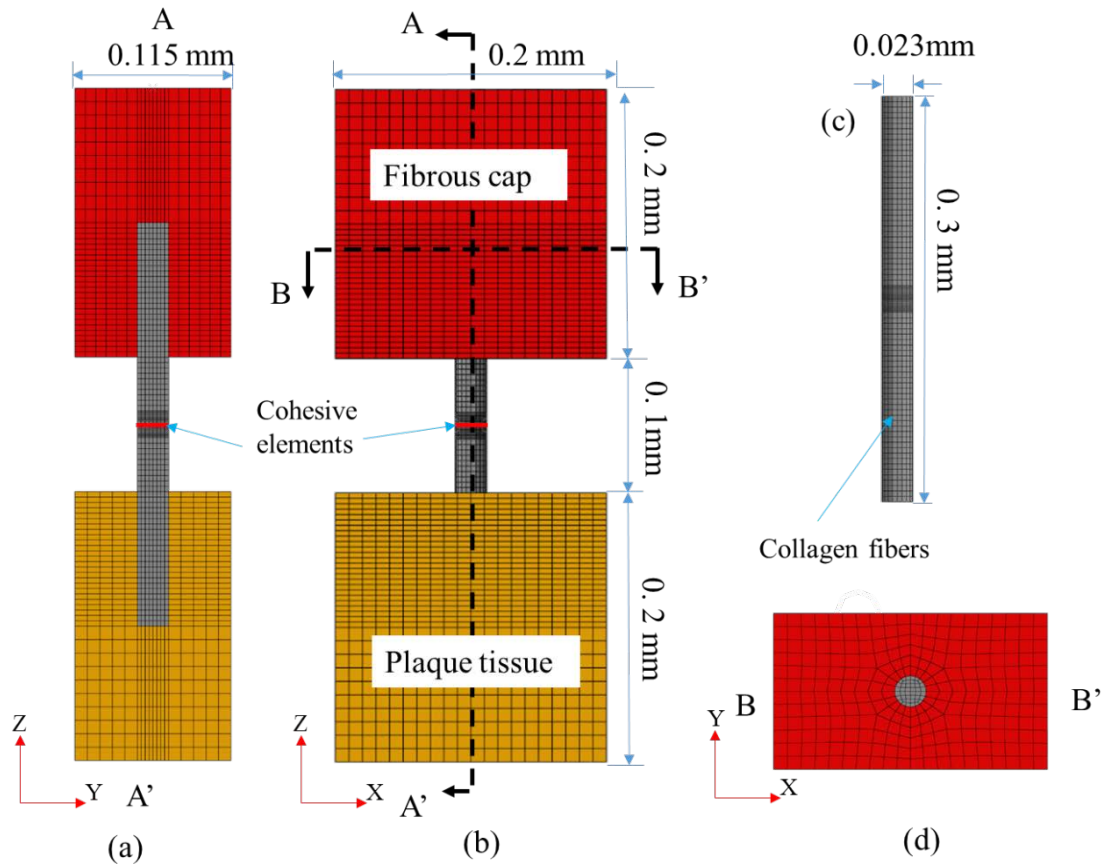

**Fig. 8.** Finite element model of the 3D unit cell for micromechanical study of fibrous cap delamination. (a) left section view of 3D unit cell, the collagen fibers connecting fibrous cap and underlying plaque tissue (the red line shows the zero thickness cohesive elements assigned to the fiber-fiber interface at the middle of fibers); (b) 2D front view of 3D unit cell; (c) collagen fibers; (d) top section view of 3D unit cell.

*Meshing*

The eight-node brick elements (C3D8H) were implemented for the embedding matrix (contains collagen fibers and smooth muscle cells, etc.). The interface between the fibers was placed with zero thickness eight-node 3D user-defined elements. The meshed geometrical model of 3D unit cell is shown in Fig. 8.

*Boundary conditions*

The unit cell had symmetrical conditions on the left and right vertical boundaries of the layers and was loaded in tension at the top and bottom boundaries by a uniform displacement. The 3D unit cell had a small volume embedded in the human fibrous cap (Fig. 7). The boundary conditions should be in accordance to the stress state in the macro model. To this end, the front and back surfaces (perpendicular to the x axial) of the fibrous cap and underlying plaque tissue of the unit cell model were constrained along x direction (Constrain the deformation along the direction of length); and the left and right surfaces (perpendicular to the y axial) were set with restriction of y direction along the width of the fibrous cap and underlying plaque tissue during delamination test. The total resultant load was determined from numerical simulation. The relation between the applied displacement and the resultant load was used to analyze the traction-separation relation of the cohesive interface between the two arterial layers.

The material parameter values of HGO model for matrix material are shown in Table 3 [23].

Table 3. Best-fit parameters of HGO model of the aorta

|  | μ (kPa) | $k_1$ (kPa) | $k_2$(-) | κ (-) | γ (degree) |
|---|---|---|---|---|---|
| Plaque | 49.45 | 23.7 | 2630 | 0.226 | 30 |
| Fibrous cap | 21.89 | 93.63 | 7957 | 0.226 | 17.22 |

The linear elastic model was used to characterize the mechanical behavior of collagen fibers, characterized as the elastic modulus and Poisson's ratio. A CZM model was adopted to describe the stiffening and softening behavior of collagen fibers during the tensile tests. The reference material parameter values of Elastic modulus, interfacial strength and critical energy release rate for the 3D unit model are set equal to the mean values of the parameters obtained from the experiment as shown in Table 2 ($G_c$=0.209 N/mm, $\sigma_c$=82.91 MPa, $E$=402.82 MPa and $\nu$=0.3).

The traction-separation curves from simulation predictions using 3D unit cell model are shown in Fig. 9. Traction was obtained through dividing the resultant force by the area of the fibrous cap-underlying plaque tissue interface (0.023 mm$^2$) and the separation was the load-point displacement. At the beginning of the traction-separation curve, the traction increased with the resistance force from the fiber-fiber interface and the matrix material. Moreover, the maximum traction occurred when the stress of fiber-fiber interface equal to the interfacial strength. At last, the traction decreased to zero when the interface of fiber-fiber damage completely.

### 6.2 Bonding strength of interface across the fibers

Collagen fibers are the major load-bearing structural constituents in the vascular tissue, which increase strength exponentially at higher strains. Hence, collagen fibers breakage is considered as the main contribution to the arterial tissue failure process. In this section, we focused on the effects of bonding strength of interface between the collagen fibers to study the traction-separation relationship of the interface between two arterial layers with fibers bridging. To gain some insight into the effect of $G_c$, five values for the $G_c$ were considered: 0.01 N/mm, 0.05 N/mm, 0.1 N/mm, 0.209 N/mm and 0.4 N/mm. Furthermore, to investigate the effect of the interfacial strength on simulation predictions, five values of $\sigma_c$ were considered: 50 MPa, 82.91 MPa, 100 MPa, 150 MPa and 200 MPa. All other aspects of the simulation model keep the same.

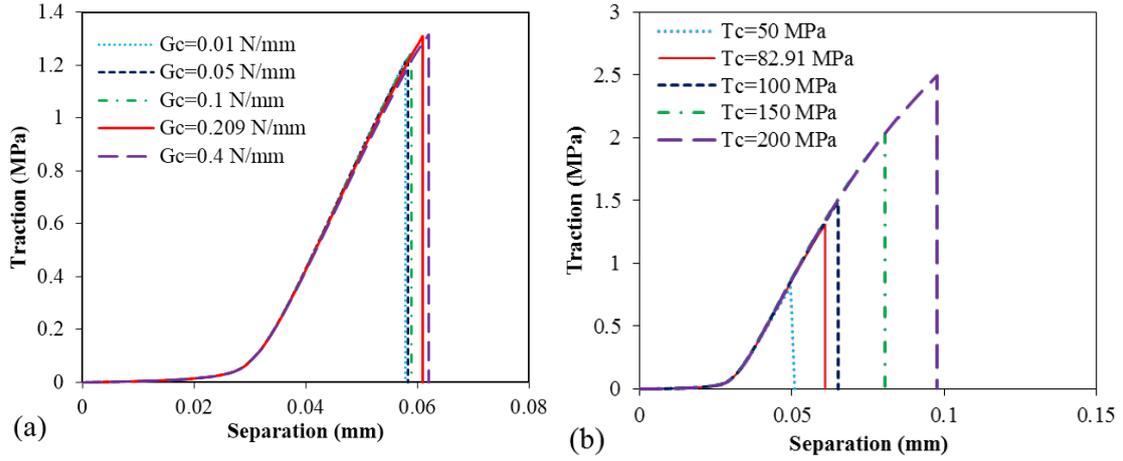

**Fig. 9.** Traction-separation curves from simulations with five sets of (a) critical energy release rate $G_c$ and (b) interfacial strength $T_c$.

The predicted traction-separation curves are shown in Fig. 9. It is seen that the traction increased while stiffness decreased with increasing critical energy release rate. The traction was largely affected by the interfacial strength that the traction increased with the increasing $T_c$, but the stiffness was not affected by the interfacial strength.

### 6.3 Variations in the fibers' stress-strain behavior

In order to gain an insight into the effect of the elastic modulus $E$ on traction-separation relation using unit cell model, five values were considered: 402.82 MPa, 500 MPa, 600 MPa, 700 MPa and 800 MPa. Furthermore, to investigate the effect of the Poisson's ratio $\nu$ on simulation predictions, five values of $\nu$ were considered: 0.1, 0.2, 0.3, 0.4 and 0.499. All other aspects of the simulation model were kept the same. Using these elastic modulus and Poisson's ratio, ten simulations were carried out (as shown in Fig. 10), each with a different $E$ or $\nu$. Other values were kept the same in all simulations.

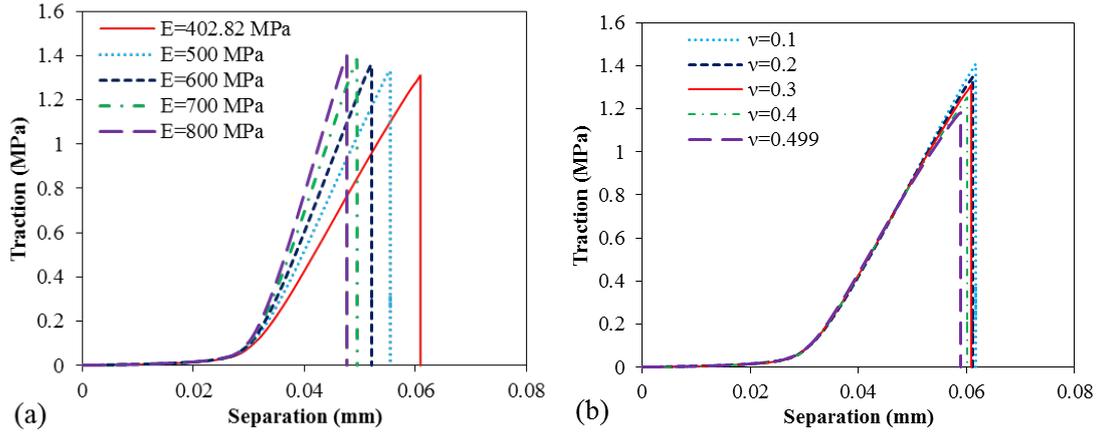

**Fig 10.** Traction-separation curves from simulations with five sets of (a) elastic modulus $E$ and (b) Poisson's ratio $\nu$.

The predicted traction-separation curves are shown in Fig. 10. It is seen that the traction and the stiffness increased with the elastic modulus. The traction and stiffness decreased with the Poisson's ratio.

**6.4 The initial gap of the interface and the length of fibers within the layers**

Considering the effect of the initial gap of the interface $l_g$ on traction-separation relation using unit cell model, five values were considered: 0 mm, 0.05 mm, 0.1 mm, 0.15 mm and 0.2 mm. All other aspects of the simulation model were kept the same.

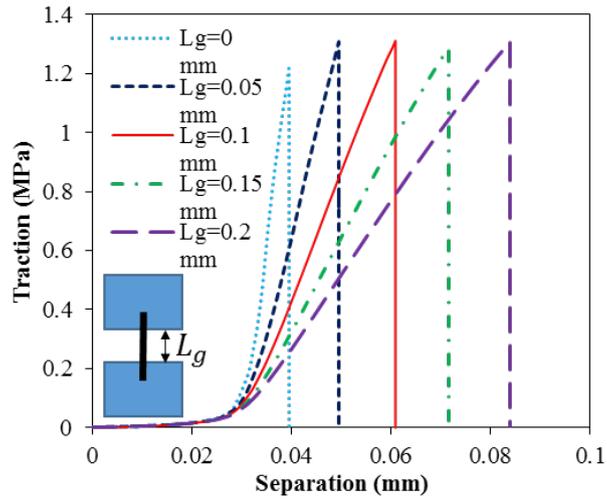

**Fig. 11.** Traction-separation curves from simulations with five sets of initial gap of the interface $L_g$.

The predicted traction-separation curves are shown in Fig. 11. It is seen that the stiffness increased with the initial gap of the interface and the predicted maximum traction was only slightly affected by the initial gap of the interface.

## 7. Discussion

During the collagen fibers breakage process, the lamellae sliding occurs and it also associated with fibrils pulling out and breakage [47]. The collagen fibers are surrounded by other collagen fibers and the matrix material such as fibroblast cells. Because some of the collagen fibers are attached to the matrix and connected to other collagen fibers (Fig. 1b), the magnitude of bonding strengths are expected to be higher than that of collagen fibers breakage. Thus, the debonding and slippage of fibers embedded within the matrix may occur, but the main contribution of the micromechanical behavior of arterial delamination is the breakage of fibers perpendicular to the delamination interface [48].

In this study, the arterial tissue contains bundles of collagen fibers (the diameter of collagen fibers ranged from 1 to 10 µm [34]). It was found that the failure strain, ultimate strength and elastic modulus of fibers were $23.00 \pm 7.3$ %, $81.75 \pm 75.14$ MPa and $402.82 \pm 334.33$ MPa, respectively. Miyazaki and Hayashi [34] showed that the mechanical properties of single collagen fibers isolated from rabbit patellar tendon and the failure strain, ultimate strength and elastic modulus were $21.6\pm3.0$ %, $8.5\pm2.6$ MPa and $54.3\pm25.1$ MPa, respectively. Except for the values of failure strain, the values obtained in this study are in agreement with those from tensile tests of single collagen fibers, the ultimate strength and elastic modulus obtained from tensile tests of bundles of collagen fibers from porcine aorta are larger than the values from Miyazaki's research. Yamamoto et al. evaluated the tensile properties of collagen fascicles (consists of collagen fibrils, fibers, interfibrillar matrix and fibroblasts [34], the diameter is approximately 300 µm [35]) from rabbit patellar tendons, and the strain at failure, ultimate strength and elastic modulus were $10.9\pm1.6$ %, $17.2\pm4.1$ MPa and $216\pm68$ MPa, respectively [35]. The values of strain at failure, ultimate strength and elastic modulus are less than those from the present study. The differences may attribute to the different tissue source of the collagen fibers, the different geometrical or structural properties of the samples.

For the identification of CZM parameter values of interface across collagen fibers, the deformation of fibers should correlate with that from tensile tests. It was noted that the fibers were stretched during the tensile test until the fibers breakage occurred at the last time point with load completely dropped to zero. Hence, the maximum effective displacement when the interface was damaged completely was very small compared to the length of fibers. The shape of the load-displacement curves is largely affected by the mechanical properties of the matrix material and the collagen fibers. Experimental results showed a nonlinearity of the mechanical response which is attributed to the nonlinear mechanical response of elastin and the gradually recruited load-bearing collagen fibers as they straighten out with increasing strain [14, 49]. At the last stage, the softening behavior of the interface across the fibers occurred and the interface damage completely and the traction dropped to zero.

For the 3D unit cell model, the parts of matrix material including fibrous cap and underlying plaque tissue were reconstructed based on the dimension of the arterial layers. The thickness, length and width should be small enough for the sake of numerical efficiency and also eliminate the effects from the boundary. For the parametric study, the material parameter values for the CZM model and elastic model of fibers were chosen within the range of the values obtained from the experiments (Table 1and Table 2) in order to simulate the actual failure process. The predicted maximum traction was largely affected by the critical energy release rate, interfacial strength and Poisson's ratio while the stiffness of the traction-separation curves was affected by the elastic modulus of fibers and the initial gap of the interface. The damage accumulated as the fibers elongate and its magnitude reached 1 when the tractions of the 3D unit cell model as well as the tractions of cohesive elements increased to the maximum values, and then the cohesive element damaged completely and the two layers of the 3D unit cell model separated.

The medicine act by decreasing the aortic tensile strength to increase the high percentage of the animal death from the aortic dissecting aneurysms [38]. It is noted that the drugs increase the ultrastructural disruption of collagen and decrease the arterial strength. For this reason, the increasing of critical energy release rate and interfacial strength of fiber-fiber interface will increase the traction of the unit cell along the failure path and the arterial strength, which will inhibit the arterial failure. Meanwhile, the effective displacement when the maximum traction of unit cell attained will increase, which will also prevent the damage of arterial tissue under certain deformation. The arterial stiffness increase with age and was taken as one factor to increase the cardiovascular disease [50]. From this study, the modulus of collagen fibers increased and the initial gap of the interface decreased, the stiffness of the interface prone to failure will increase, which will accelerate the damage of the interface under a certain small deformation. Therefore, the parametric study using 3D unit cell provides a possibility to investigate the mechanism of drug treatments to the arterial tissue failure.

Comparisons of simulation predictions of the load-displacement curve with experimental measurements revealed that the simulation predictions were able to capture the essential features of the load-displacement curve from the collagen fibers tensile tests and showed a good quantitative agreement with the experimental measurements. The results from the simulation predictions of fibers breakage validated the proposed CZM based approach for modeling and simulating collagen fibers breakage events. Furthermore, the parametric study using 3D unit cell model provides a method to investigate the traction-separation relationship of fibers bridging across the arterial layers at the micromechanical scale.

Despite the novelty of some experimental observations and the encouraging predictive power of the proposed micromechanical model of the arterial tissue, certain limitations should be considered in the interpretation of the obtained results. Firstly, the fibers were assumed with constant cross-section area along the longitudinal direction, in actuality the area of cross-section varies with irregular shape. Moreover, the fibers are composed of bundles of collagen fibers twisted together. Secondly, the cross-section of fibers was assumed to be a smooth interface perpendicular to the axial direction in the simulation of fibers breakage process. But, the breakage area may not be an ideal cross-section because the fibrils pull-out and breakage occurs inside the fibers. Future studies should incorporate these factors in the development of the modeling.

## 8. Conclusions

In both clinical and pharmaceutical researches, including disease diagnosis and medicine development, it is advantageous to understand the arterial mechanical response under failure states. For the vascular biomechanics, it is necessary to determine the appropriate theoretical model which is able to describe tissue mechanical properties and help to formulate and solve specific boundary value problems. In the current study, a 3D unit cell model was developed and applied successfully to do the parametric study of the arterial tissue failure process at the microscopic scale. Quantifying the failure mechanical response of arterial tissue is necessary to understand its pathophysiological mechanical performance, and may shed light on the genesis and progression of abdominal aortic aneurysm delamination.

**ACKNOWLEDGEMENTS**

The authors gratefully acknowledge the sponsorship of NSF (award # CMMI-1200358) and this work was partially supported by a SPARC Graduate Research Grant from the Office of the Vice President for Research at the University of South Carolina.